\documentclass[a4paper]{PoS}

\usepackage{textcomp}
\usepackage{amsfonts} % AMS
\usepackage{amssymb} % AMS
\usepackage{amsmath} % AMS
\usepackage{slashed}
\usepackage{graphicx}
\usepackage{pifont}
\usepackage{multirow,lscape}
\usepackage{array}
\usepackage{longtable}
\usepackage[FIGBOTCAP,TABBOTCAP,tight]{subfigure}
\usepackage{verbatim}
\usepackage{bm}
\usepackage{comment}
\usepackage{xcolor}
\usepackage{url} %%% URL

\graphicspath{{./figs/}}

\newcommand{\wtd}[1] { \widetilde{#1} }

\newcommand{\GeV}{\mathop{\rm GeV}\nolimits}

\newcommand{\MSb}{\mathop{\rm \overline{MS}}\nolimits}

\title{Non-perturbative Renormalization of Bilinear Operators on Fine Lattice}

\ShortTitle{NPR on Fine Lattice}

\author{
Hwancheol Jeong, 
Weonjong Lee, 
Jeonghwan Pak, 
Sungwoo Park
\\
Lattice Gauge Theory Research Center, CTP, and FPRD, \\
Department of Physics and Astronomy, \\
Seoul National University, Seoul, 151-747, South Korea\\
E-mail: \email{wlee@snu.ac.kr}}

\author{
\speaker{Jangho Kim} 
\\
National Institute of Supercomputing and Networking, \\
Korea Institute of Science and Technology Information, Daejeon 34141, Korea\\
E-mail: \email{jangho@kisti.re.kr}
}

\author{SWME Collaboration}

\abstract{
We present results of the wave function renormalization factor $Z_q$
and mass renormalization factor $Z_m$ obtained using non-perturbative
renormalization (NPR) method in the RI-MOM scheme with HYP improved
staggered quarks.
We use fine ensembles of MILC asqtad lattices ($N_f = 2+1$) with $28^3
\times 96$ geometry, $a \approx 0.09$\,fm, and $am_\ell/am_s =
0.0062/0.031 $.
We also study on scalability of $Z_q$ and $Z_m$ by comparing the
results on the coarse and fine ensembles.
}

\FullConference{The 33rd International Symposium on Lattice Field Theory\\
                 14 -18 July  2015\\
                 Kobe International Conference Center, Kobe, Japan}

\begin{document}
%-------------
% Introduction
%-------------
\section{Introduction}
In our previous work~\cite{Kim:2013bta}, we presented the results of
the wave function renormalization factor $Z_q$, mass renormalization
factor $Z_m$ and the complete set of renormalization factors for
bilinear operators obtained on the $20^3 \times 64$ MILC asqtad coarse
lattice at $a \approx 0.12$\,fm with $am_{\ell}/am_s=0.01/0.05$.
In this proceeding, we analyse the $Z_q$ and $Z_m$ on the
$28^3\times96$ MILC asqtad fine lattices ($a\approx 0.09$\,fm,
$am_\ell/am_s=0.0062/0.031$) and compare the results with those on the
coarse lattices.
%

%--------
% Results
%--------
\section{Results}
We calculate the renormalization factors with Landau gauge fixing
using HYP-smeared staggered quarks.
To do the chiral extrapolation, we perform the measurements with 5
valence quark masses ($am = 0.0062, 0.0124, 0.0186, 0.0248, 0.031$)
on the MILC fine ensembles at $a\approx 0.09$\,fm.
We also carry out the measurements for 20 external momenta given
in Table \ref{tab:momentum}.
The measurements are done over 30 gauge configurations.
%
%---------------------
% momentum list table
%---------------------
%
\begin{table}[h!]
\begin{small}
\centering
\begin{tabular}{c | c | c || c | c | c || c | c | c }
\hline
\hline
$n(x, y, z, t)$ & $a|\wtd{p}|$ & GeV &
$n(x, y, z, t)$ & $a|\wtd{p}|$ & GeV &
$n(x, y, z, t)$ & $a|\wtd{p}|$ & GeV \\
\hline
$(1,1,1,3)$  & 0.4355 & 1.0197 & 
$(1,1,1,4)$  & 0.4686 & 1.0974 & 
$(1,2,1,4)$  & 0.6088 & 1.4257 \\
$(1,2,1,6)$  & 0.6755 & 1.5819 & 
$(2,1,2,6)$  & 0.7794 & 1.8250 &
$(2,2,2,7)$  & 0.9023 & 2.1130 \\
$(2,2,2,8)$  & 0.9372 & 2.1947 & 
$(2,2,2,9)$  & 0.9753 & 2.2839 & 
$(2,3,2,7)$  & 1.0324 & 2.4177 \\
$(2,3,2,8)$  & 1.0631 & 2.4895 & 
$(2,3,2,9)$  & 1.0968 & 2.5684 &
$(3,2,3,8)$  & 1.1756 & 2.7529 \\
$(3,3,3,7)$  & 1.2528 & 2.9337 & 
$(3,3,3,8)$  & 1.2782 & 2.9931 &
$(3,3,3,10)$ & 1.3371 & 3.1312 \\
$(3,4,3,9)$  & 1.4349 & 3.3602 &
$(4,3,4,10)$ & 1.5789 & 3.6973 &
$(4,4,4,10)$ & 1.6868 & 3.9501 \\
$(4,4,4,12)$ & 1.7418 & 4.0788 &
$(4,4,4,14)$ & 1.8046 & 4.2259 &
            &        &        \\
\hline
\hline
\end{tabular}
\caption{
\label{tab:momentum}
The list of momenta used for our analysis.
The first column is the four vectors in the units of
$(\dfrac{2\pi}{L_s}, \dfrac{2\pi}{L_s}, \dfrac{2\pi}{L_s},
\dfrac{2\pi}{L_t})$, where $L_s$ ($L_t$) is the number of sites in
the spatial (temporal) direction.
}
\end{small}
\end{table}

\subsection{Wave Function Renormalization Factor $Z_q$}
\begin{figure}[tbhp]
\centering
\includegraphics[width=0.6\columnwidth]{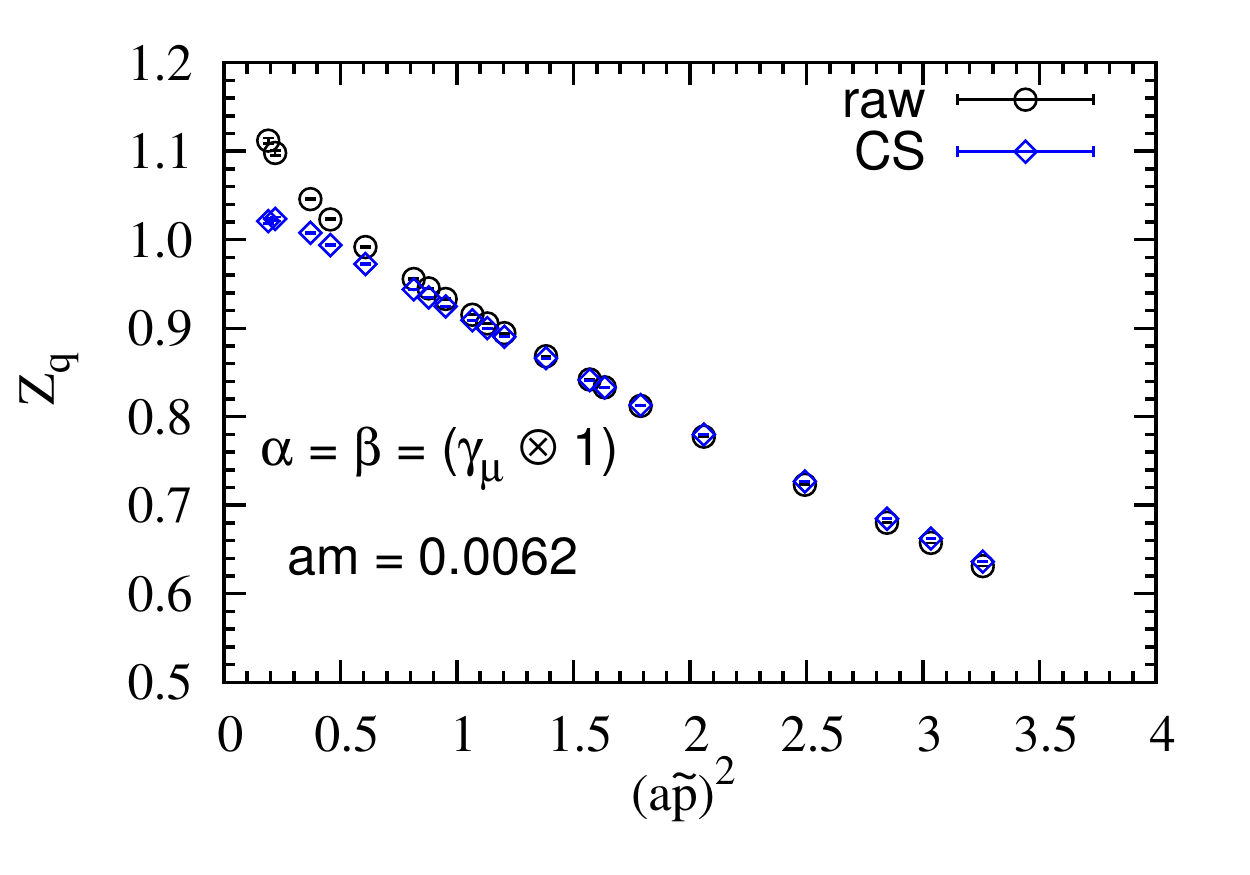}
\caption{ $Z_q$ obtained from conserved vector current ($V \times S$)
  at a fixed quark mass $(am = 0.0062)$.  The black circles represent
  raw data and blue diamonds are CS data at a CS $\mu_0=3\GeV$.}
\label{fig:Z_q:CS}
\end{figure}

\begin{figure}[t!]
\subfigure[m-fit for $Z_q$]{
\includegraphics[width=0.48\textwidth]{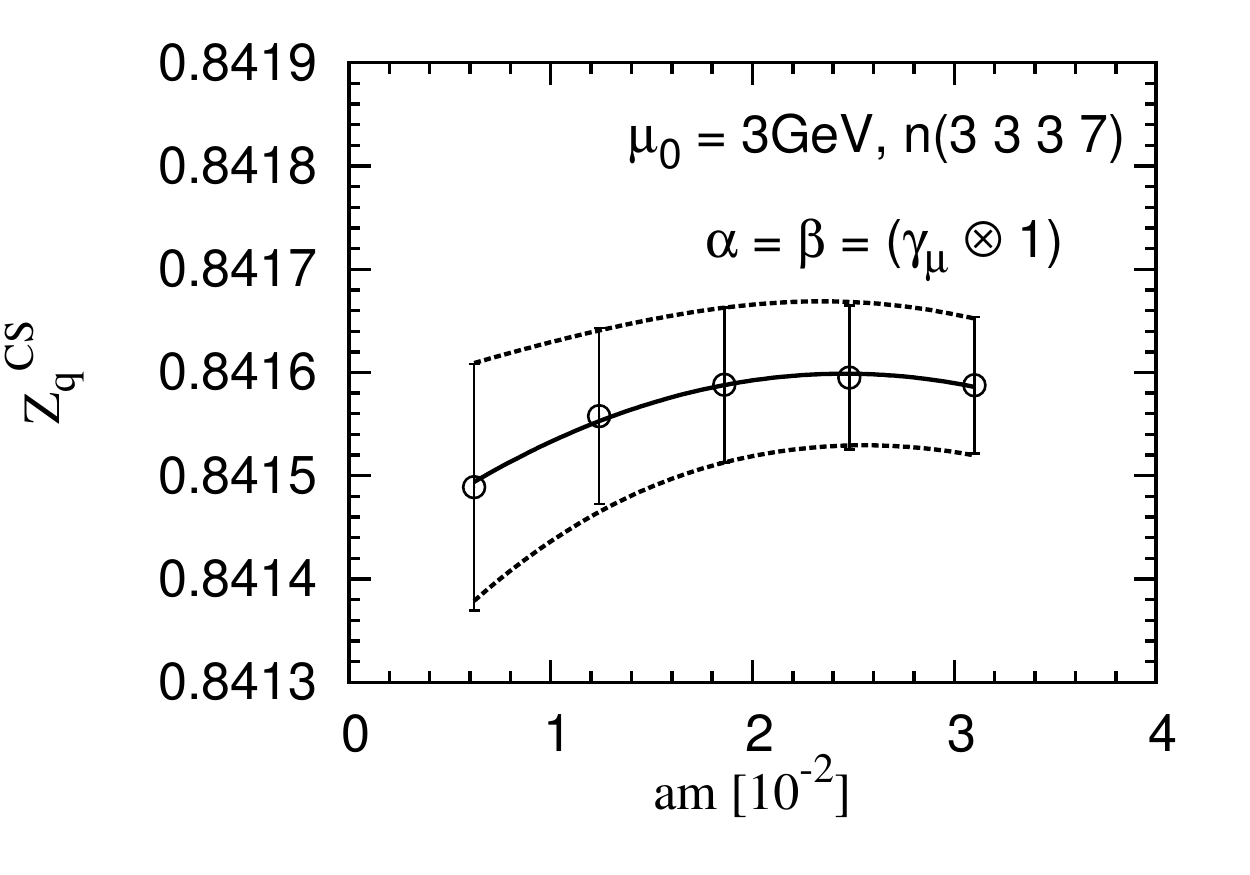}
\label{sfig:Z_q:m-fit}
}
\hfill
\subfigure[p-fit for $Z_q$]{
\centering
\includegraphics[width=0.48\textwidth]{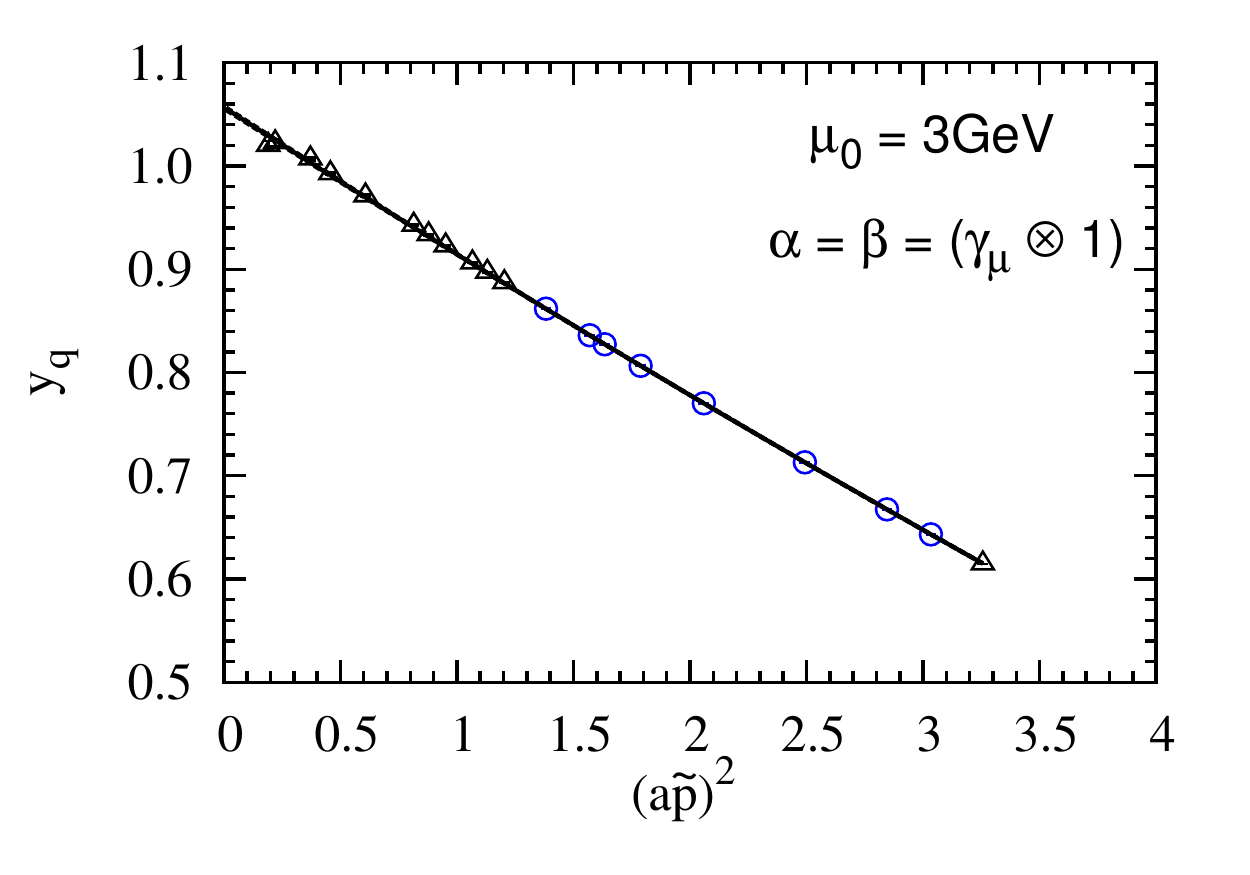}
\label{sfig:Z_q:p-fit}
}
\caption{ \protect\subref{sfig:Z_q:m-fit} m-fit results for $Z_q$ at a reduced
  momentum $n=(3,3,3,7)$ and \protect\subref{sfig:Z_q:p-fit} p-fit results for
  $y_q$.  Here, we use the conserved vector current at $\mu_0=3\GeV$.
  $y_q \equiv Z_q(\mu_0,am=0) - \langle c_4 \rangle (a\wtd{p})^4$.
  The blue circles are used for fitting.  }
\label{fig:VxS}
\end{figure}

Let us consider the conserved vector current to obtain the wave function
renormalization factor $Z_q$.
We use the same method as in Ref.~\cite{Kim:2013bta} to obtain the
$Z_q$.
First, we convert the raw data to the data defined at a common scale
(CS) $\mu_0 = 3$\,GeV using the four-loop RG evolution equation in
Ref.~\cite{Chetyrkin:1999pq, Aoki:2007xm}.
In Fig.~\ref{fig:Z_q:CS}, we present the raw data as the black circles
and CS data as blue diamonds as a function of the square of reduced
momentum $(a\wtd{p})^2$ at a fixed quark mass $(am = 0.0062)$.
After converting the raw date to the CS data, we perform the fitting
with respect to quark masses at a fixed external momentum to the
following fitting function. We call this m-fit.
\begin{align}
f_{\text{m-fit}} = b_1 + b_2 \cdot am + b_3 \cdot (am)^2
\end{align}
The fitting results are presented in Table~\ref{tab:Z_q:m-fit} and the plot is
given in Fig.~\ref{sfig:Z_q:m-fit}.
\begin{table}[htbp]
\center
\begin{tabular}{c | c | c || c }
\hline
\hline
$b_1$ & $b_2$ & $b_3$ & $\chi^2/\text{dof}$ \\
\hline
0.84141(15) & 0.0153(97) & -0.31(17) & 0.004(10) \\
\hline
\hline
\end{tabular}
\caption{ m-fit results for $Z_q$ at $\mu_0 = 3\GeV$ for a fixed
  external momentum $n=(3,3,3,7)$.  }
\label{tab:Z_q:m-fit}
\end{table}

We take $b_1$ as the chiral limit values which are function of
external momentum $(a\wtd{p})^2$.
After m-fit, we fit $b_1$ to the following fitting function. We call
this p-fit.
\begin{align}
\label{eq:Z_q:p-fit}
f_{\text{p-fit}} = c_1 + c_2 (a\wtd{p})^2
+ c_3 \cdot ((a\wtd{p})^2)^2 + c_4 \cdot (a\wtd{p})^4
\end{align}
The fitting results are presented in Table~\ref{tab:Z_q:p-fit} and the
plot is presented in Fig.~\ref{sfig:Z_q:p-fit}.
\begin{table}[h!]
\center
\begin{tabular}{ c | c | c | c || c }
\hline
\hline
$c_1$ & $c_2$ & $c_3$ & $c_4$ & $\chi^2/\text{dof}$ \\
\hline
1.0567(11) & -0.1452(10) & 0.00294(14) & 0.0082(11) & 0.13(26) \\
\hline
\hline
\end{tabular}
\caption{ P-fit results for $Z_q$ at $\mu_0 = 3\GeV$.}
\label{tab:Z_q:p-fit}
\end{table}
The $\mathcal{O}((a\wtd{p})^2)$ and higher order terms correspond to
lattice artifacts.
Hence, we take $c_1$ as $Z_q$ value in RI-MOM scheme at $\mu_0 =
3\GeV$.
Using the four-loop RG running formula~\cite{Chetyrkin:1999pq,
  Aoki:2007xm}, we convert the $Z_q$ from the RI-MOM scheme to the
$\MSb$ scheme.

We estimate the systematic error in two different ways. 
One systematic error comes from truncation of four-loop RG running
factor which is used to convert the $Z_q$ from the RI-MOM scheme to
the $\MSb$ scheme.
Hence, we take five-loop uncertainty ($\sim\mathcal{O}(\alpha_s^4)$)
and define $E_t$ as follows.
\begin{align}
E_t = Z_q^{\text{RI-MOM}} \cdot (\alpha_s)^4
\end{align}
The other systematic error comes from the difference between the
conserved vector and axial currents.
Theoretically, $Z_q$ obtained from the conserved vector and axial
currents must be identical to each other. However, they are not same
in our study. Hence, we take the difference of them as the systematic
error and define $E_{\Delta}$ as follows.
\begin{align}
E_{\Delta} = |Z_q(V \otimes S) - Z_q(A \otimes P)|
\end{align}
The total error ($E_\text{tot}$) is obtained adding the statistical
error ($E_\text{stat}$) and the systematic errors in quadrature.
We present the final result of $Z_q$ in $\MSb$ scheme at $\mu_0 =
3\GeV$ and its statistical and systematic errors in Table
\ref{tab:Z_q:MSb}.
\begin{table}[htbp]
\center
\begin{tabular}{ c || c | c | c | c }
\hline
\hline
$Z_q^{\MSb}(\mu_0)$ & $E_\text{stat}$ & $E_t$ & $E_{\Delta}$
& $E_\text{tot}$ \\
\hline
1.0494 & 0.0011 & 0.0038 & 0.0099 & 0.0107 \\
\hline
\hline
\end{tabular}
\caption{ $Z_q$ in the $\MSb$ scheme at $\mu_0=3\GeV$ with statistical
  and systematic errors. }
\label{tab:Z_q:MSb}
\end{table}

\subsection{Quark Mass Renormalization Factor $Z_m$}
\begin{figure}[t!]
\centering
\includegraphics[width=0.6\columnwidth]{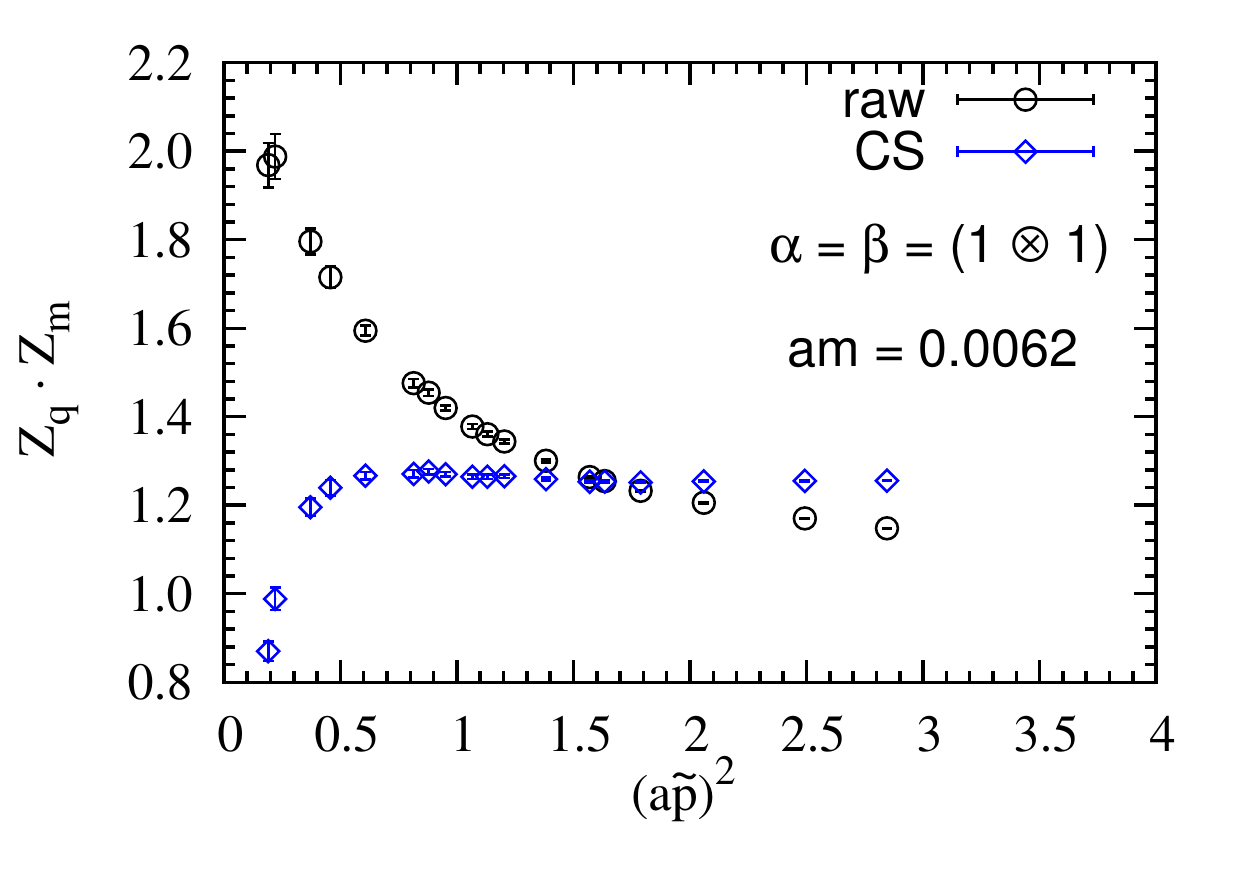}
\caption{ $Z_q \cdot Z_m$ obtained from $[S \times S]$ bilinear
  operator at $am = 0.0062$. Here, $\mu_0 = 3$\,GeV.}
\label{fig:Z_m:CS}
\end{figure}

%---------------------
% Mass Renormalization
%---------------------
%
Quark mass renormalization factor $Z_m$ is obtained from the bilinear
operator $[S \otimes S]$.
Here, we use the same analysis method as in Ref.~\cite{Kim:2013bta}.
Note that we analyse $Z_q \cdot Z_m$ instead of $Z_m$ directly. 
After we obtain the $Z_q \cdot Z_m$ in RI-MOM scheme at $\mu_0=3\GeV$
through m-fit and p-fit, we divide by $Z_q$ obtained from the
conserved vector current.
First, we convert raw data to the CS data using the four-loop RG
running formula for $Z_q \cdot Z_m$.
We present the raw and CS data for $Z_q \cdot Z_m$ in
Fig.~\ref{fig:Z_m:CS}.

\begin{figure}[t!]
\subfigure[m-fit]{
\includegraphics[width=0.48\textwidth]{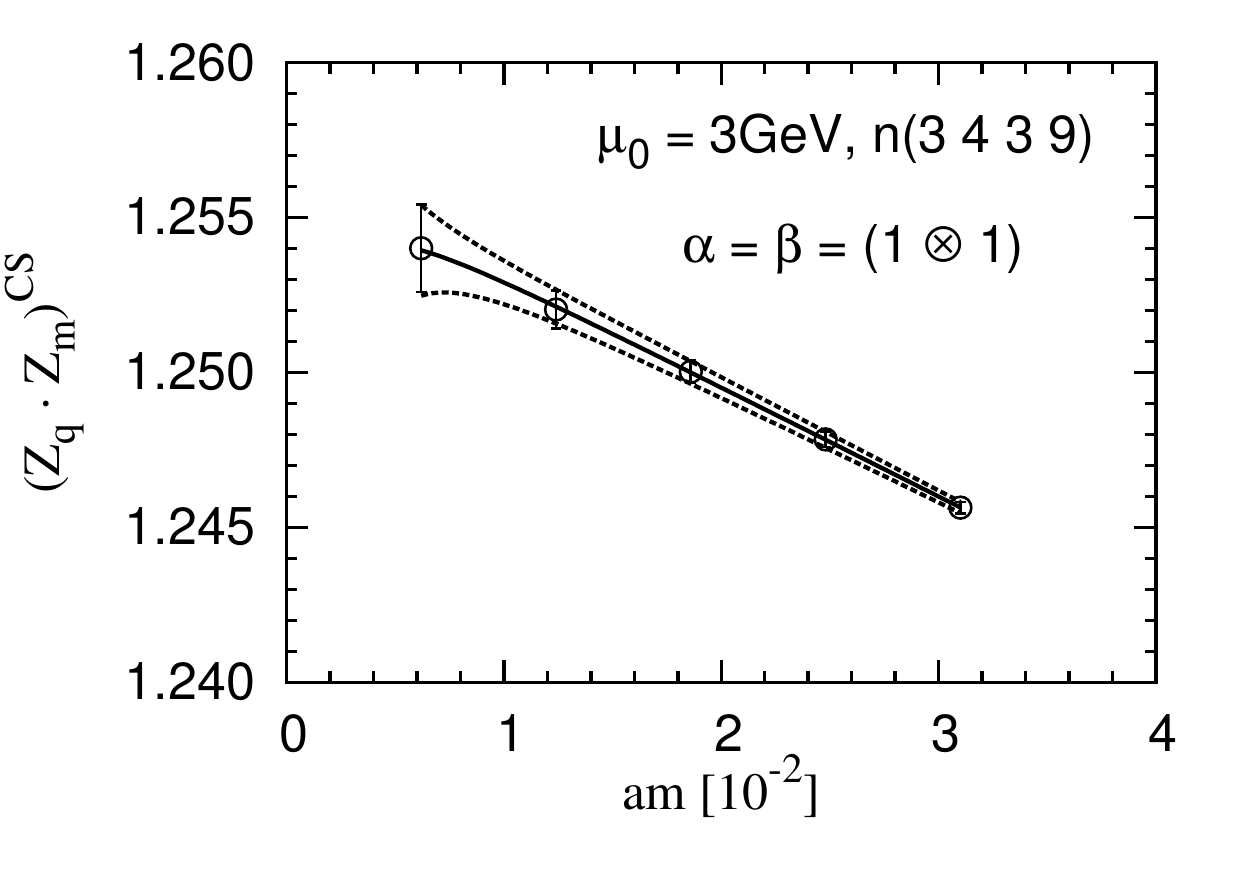}
\label{sfig:SxS_mass}
}
\hfill
\subfigure[p-fit]{
\centering
\includegraphics[width=0.48\textwidth]{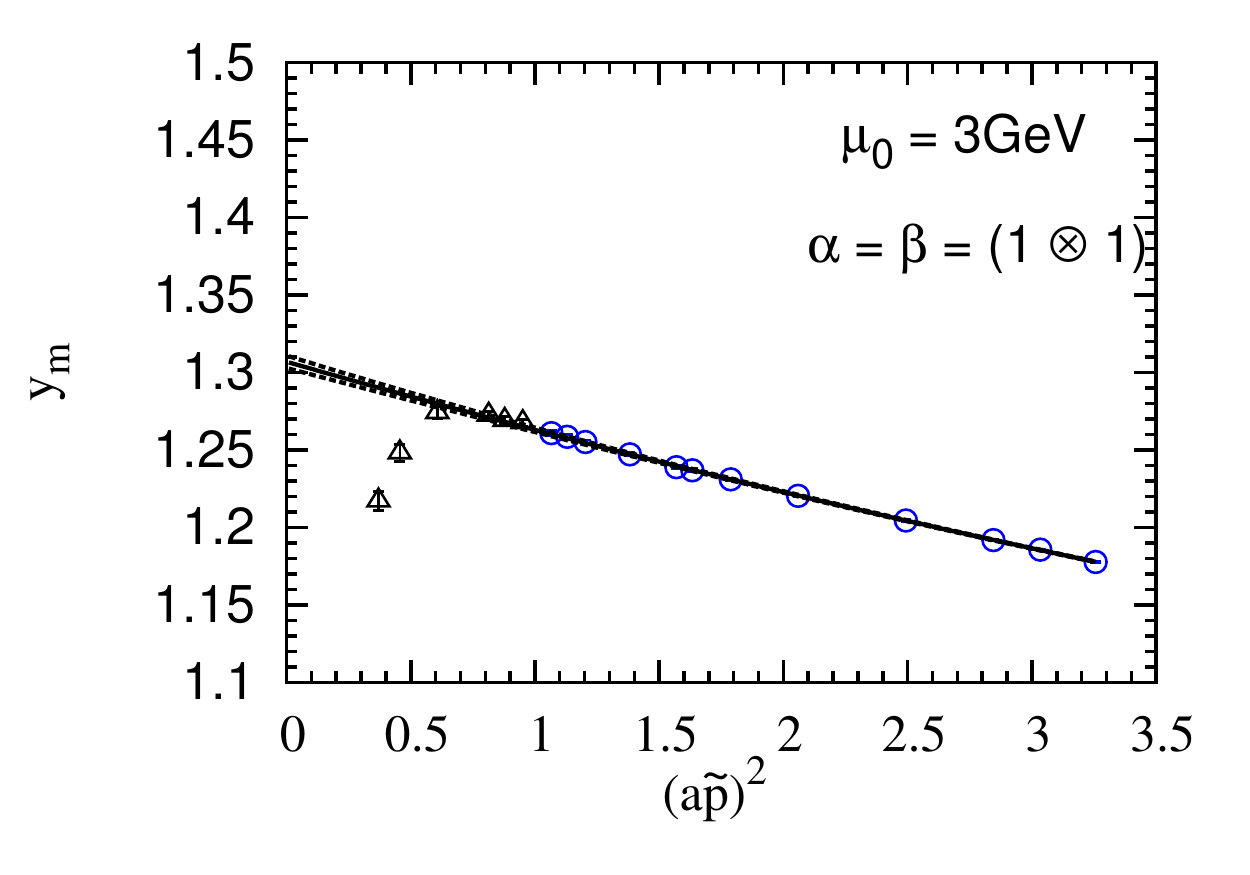}
\label{sfig:SxS_mom}
}
\caption{ Fitting results of $Z_q \cdot Z_m$ for
  \protect\subref{sfig:SxS_mass} m-fit and \protect\subref{sfig:SxS_mom} p-fit.  For
  the m-fit, the reduced momentum is fixed to $n=(3,4,3,9)$.  For the
  p-fit, $y_m \equiv (Z_q \cdot Z_m)(\mu_0, am=0) - \langle h_4
  \rangle (a\wtd{p})^4$.  The blue circle data are used for fitting.
}
\label{fig:Z_m}
\end{figure}

Using the CS data for $Z_q \cdot Z_m$, we carry out m-fit and p-fit.
The fitting functions for m-fit and p-fit are
\begin{align}
g_{\text{m-fit}} &= d_1 + d_2 \cdot am + d_3 \cdot \frac{1}{(am)^2}\\
\label{eq:Z_m:p-fit}
g_{\text{p-fit}} &= h_1 + h_2 (a\wtd{p})^2 + h_3 \cdot ((a\wtd{p})^2)^2 + h_4 \cdot (a\wtd{p})^4
\,.
\end{align}
We present fitting results of m-fit in Table~\ref{tab:Z_m:m-fit}
and in Fig.~\ref{fig:Z_m}\;\subref{sfig:SxS_mass}.
We show fitting results of p-fit in Table~\ref{tab:Z_m:p-fit} and in
Fig.~\ref{fig:Z_m}\;\subref{sfig:SxS_mom}.

\begin{table}[htbp]
\center
\begin{tabular}{c | c | c || c }
\hline
\hline
$d_1$ & $d_2$ & $d_3$ & $\chi^2/\text{dof}$ \\
\hline
1.25664(60) & -0.354(15) & -0.000000019(51) & 0.017(25) \\
\hline
\hline
\end{tabular}
\caption{ Fitting results of $Z_q \cdot Z_m$ for m-fit.  The reduced
  momentum is fixed to $n=(3,4,3,9)$.  }
\label{tab:Z_m:m-fit}
\end{table}
\begin{table}[htbp]
\center
\begin{tabular}{c | c | c | c || c }
\hline
\hline
$h_1$ & $h_2$ & $h_3$ & $h_4$ & $\chi^2/\text{dof}$ \\
\hline
1.3069(39) & -0.0459(35) & 0.00191(62) & 0.0308(45) & 0.37(32) \\
\hline
\hline
\end{tabular}
\caption{ Fitting results of $Z_q \cdot Z_m$ for p-fit.  }
\label{tab:Z_m:p-fit}
\end{table}

\begin{table}[h!]
\center
\begin{tabular}{c || c | c | c | c }
\hline
\hline
$Z_m^{\MSb}(\mu_0)$ & $E_{stat}$ & $E_t$ & $E_{\Delta}$ & $E_{tot}$ \\
\hline
1.0117 & 0.0032 & 0.0044 & 0.0005 & 0.0055 \\
\hline
\hline
\end{tabular}
\caption{ $Z_m$ in $\MSb$ scheme at $\mu_0=3\GeV$. }
\label{tab:Z_m:MSb}
\end{table}

We determine $Z_m$ by dividing $Z_q \cdot Z_m$ by $Z_q$ obtained using
the conserved vector current.
Then, we convert $Z_m$ in the RI-MOM scheme into that in the $\MSb$ scheme
using the four-loop RG evolution formula.
\begin{align}
  Z_m^{\MSb}(\mu_0) = U(\infty \to \mu_0, \MSb) \;
  U(\mu_0 \to \infty, \text{RI-MOM}) \;
  Z_m^\text{RI-MOM} (\mu_0) \,,
\end{align}
where $U(\mu_1 \to \mu_2, R)$ is the RG evolution matrix from the scale
$\mu_1$ to $\mu_2$ in the $R$ scheme.
The results are summarized in Table \ref{tab:Z_m:MSb}.
Here, the systematic errors are estimated in the same way as in $Z_q$.

\subsection{Comparison of $Z_q$ and $Z_m$ between coarse and fine lattices}
\begin{figure}[t!]
\subfigure[p-fit]{
\includegraphics[width=0.48\textwidth]{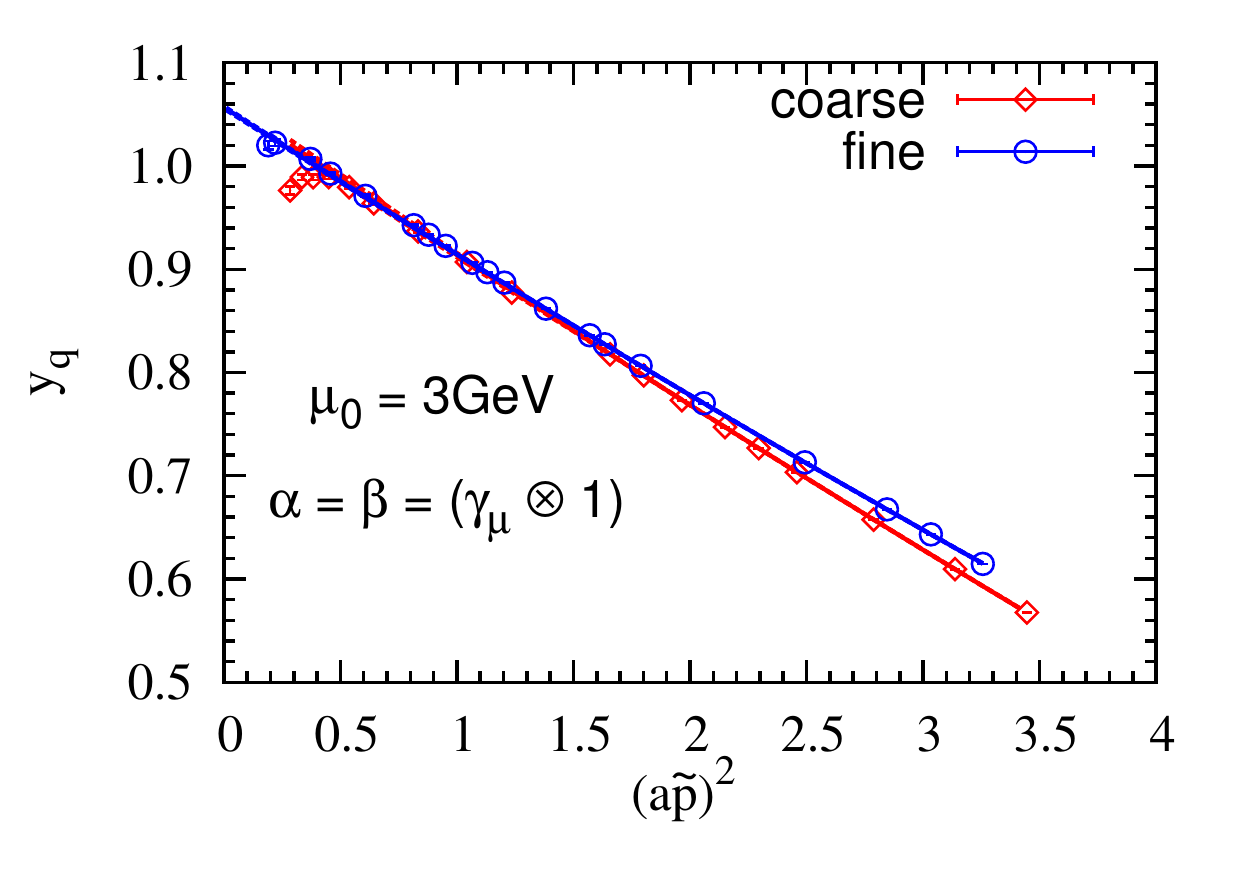}
\label{sfig:Zq_mom_coarse_fine}
}
\hfill
\subfigure[final results]{
\centering
\includegraphics[width=0.48\textwidth]{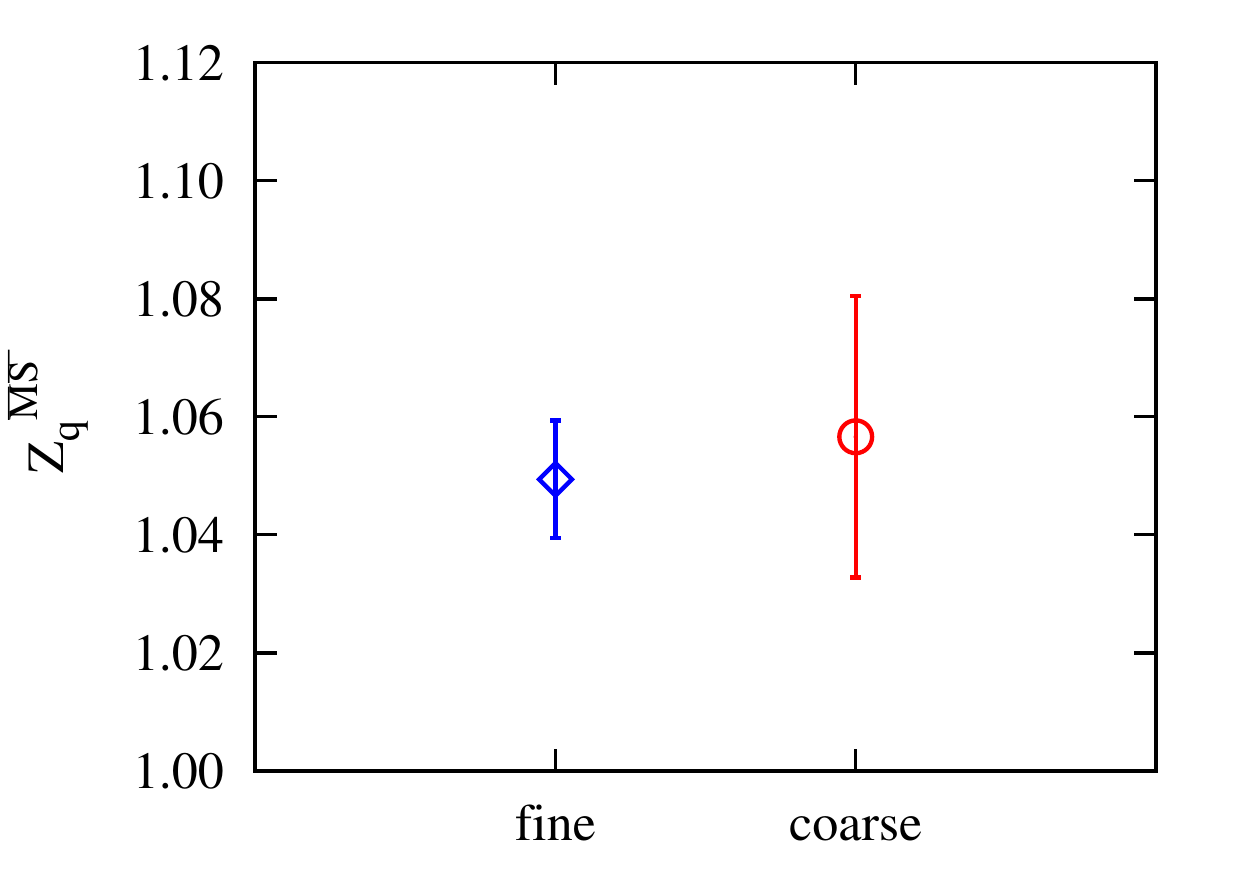}
\label{sfig:Zq_coarse_fine}
}
\caption{ Comparison of $Z_q(\mu_0)$ on the coarse and fine lattices.
  The red (blue) data represent results on the coarse (fine) lattice.
}
\label{fig:Zq_coarse_fine}
\end{figure}

\begin{figure}[t!]
\subfigure[p-fit]{
\includegraphics[width=0.48\textwidth]{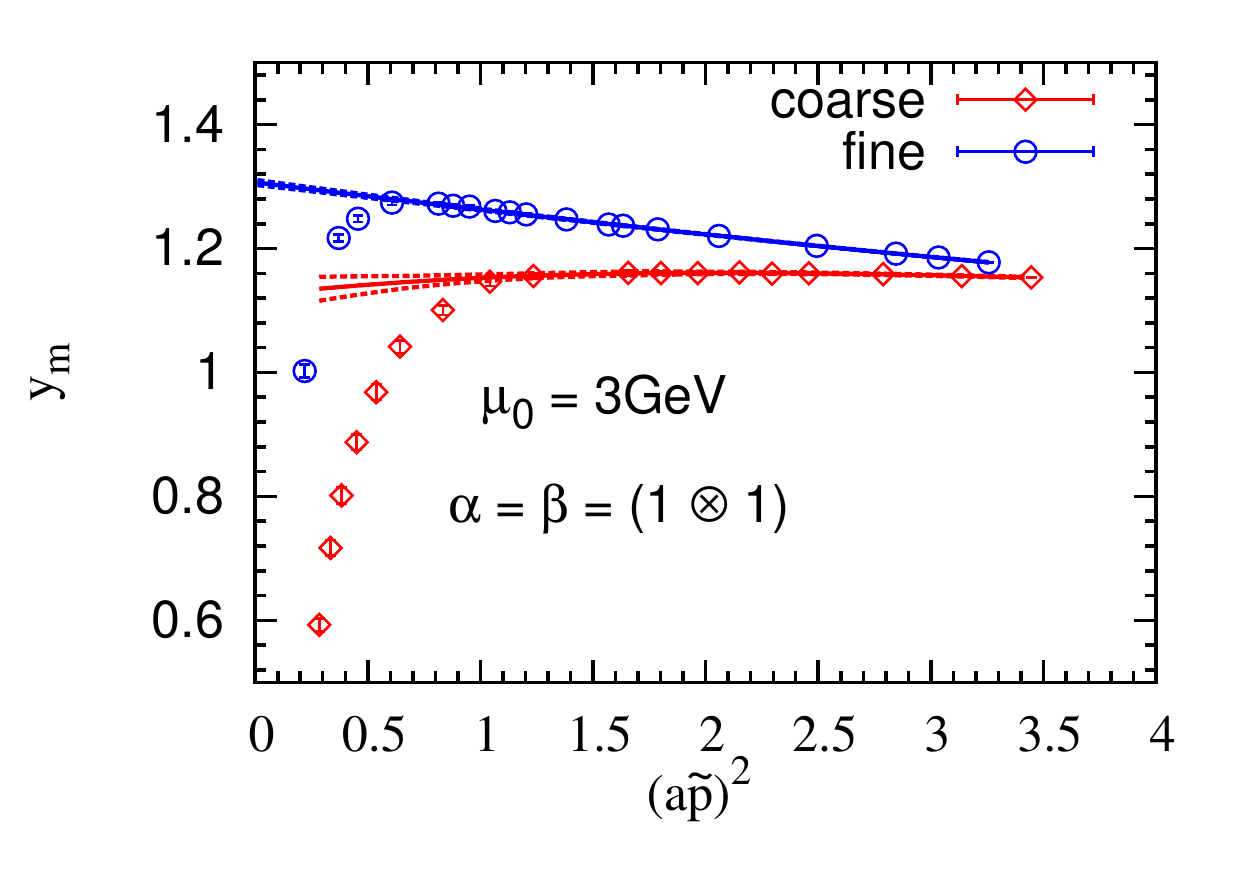}
\label{sfig:Zm_mom_coarse_fine}
}
\hfill
\subfigure[final results]{
\centering
\includegraphics[width=0.48\textwidth]{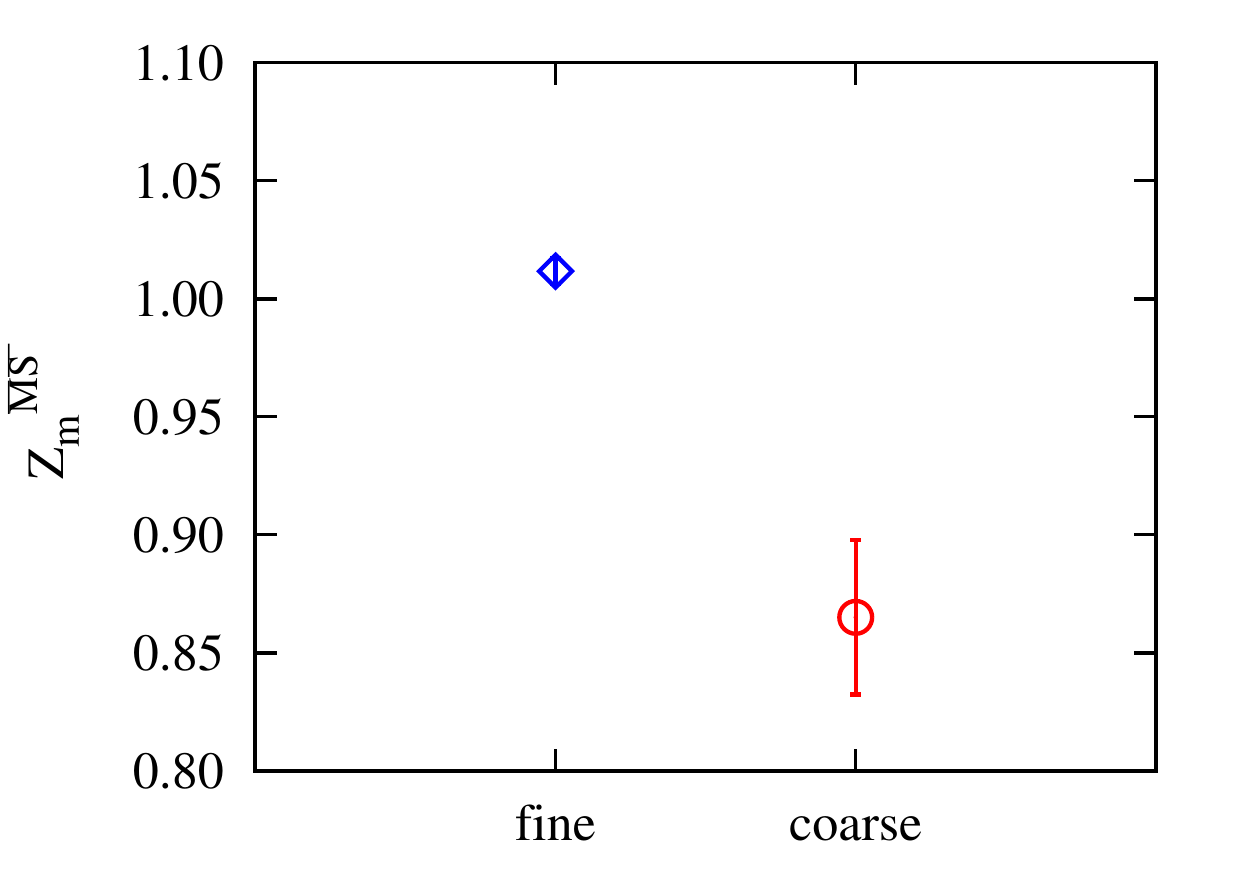}
\label{sfig:Zm_coarse_fine}
}
\caption{ Comparison of $Z_m(\mu_0)$ on the coarse and fine lattices.
  The red (blue) data represent results on the coarse (fine) lattice.
}
\label{fig:Zm_coarse_fine}
\end{figure}

Now, we present the comparison of the $Z_q$ and $Z_m$ results at $\mu_0=3\GeV$
between coarse and fine lattices in Fig.~\ref{fig:Zq_coarse_fine} and
Fig.~\ref{fig:Zm_coarse_fine}.  

The results of $Z_q$ and $Z_m$ at $\mu_0=3\GeV$ on coarse and fine
lattices are presented in Table~\ref{tab:coarse_fine}.
We find that the total errors of $Z_q^{\MSb}$ and $Z_m^{\MSb}$ on fine
lattice are reduced dramatically compared with those of coarse
lattice.
\begin{table}[htbp]
\center
\begin{tabular}{c || c | c }
%
%\hline
\hline
 & coarse & fine \\
\hline\hline
$Z_q^{\MSb}$(3GeV) & 1.0566(59)(231) & 1.0494(11)(99) \\
\hline
$Z_m^{\MSb}$(3GeV) & 0.865(21)(25)   & 1.0117(32)(44)  \\
\hline
%\hline
%
\end{tabular}
\caption{ The comparison of $Z_q$ and $Z_m$ at $\mu_0 = 3\GeV$ between
  coarse and fine lattices.  The first error is statistical and the
  second is systematic. }
\label{tab:coarse_fine}
\end{table}
%

%%%{\color{red} \textbf{EDIT-wlee}}

\section{Conclusion}
Here, we present the results of the wave function renormalization
factor $Z_q$ and mass renormalization factor $Z_m$ for the staggered
bilinear operators defined in the $\MSb$ scheme at $\mu_0 = 3$\;GeV.
We use the NPR method in the RI-MOM scheme as an intermediate scheme.
We use one of the MILC asqtad fine ($a\approx 0.09$\;fm) ensembles to
calculate the matching factors.
By comparing results with those on the coarse ensembles, we find that
the statistical and systematic errors of $Z_q$ and $Z_m$ are reduced
dramatically on the fine lattice.
We plan to extend the calculation to the superfine ($a\approx
0.06$\;fm) and ultrafine ($a\approx 0.045$\;fm) ensembles in the
future.
As a consequence, we will study on the scalability of $Z_q$ and $Z_m$.
We also plan to calculate the renormalization factors on the fine
ensembles with different sea quark masses, which will help us to
understand their dependence on sea quark masses.

\acknowledgments
J.~Kim is supported by Young Scientists Fellowship through National
Research Council of Science \& Technology (NST) of KOREA.
The research of W.~Lee is supported by the Creative Research
Initiatives Program (No.~2015001776) of the NRF grant funded by the
Korean government (MEST).
W.~Lee would like to acknowledge the support from the KISTI
supercomputing center through the strategic support program for the
supercomputing application research (No.~KSC-2014-G3-002) with much
gratitude.
Computations were carried out on the DAVID GPU clusters at Seoul
National University.

\bibliography{refs}

\end{document}